\documentclass[conference]{IEEEtran}
\IEEEoverridecommandlockouts
\usepackage{amsmath,amssymb,amsfonts}
\usepackage{algorithmic}
\usepackage{graphicx}
\usepackage{textcomp}
\usepackage[numbers,sort]{natbib}
\usepackage{multirow}
\usepackage{xcolor}
\usepackage{nameref}
\usepackage[normalem]{ulem}
\usepackage{listings}
\usepackage{xcolor}

\usepackage[compatibility=false]{caption}
\usepackage{comment}
\usepackage{enumitem}
\usepackage[utf8]{inputenc}
\DeclareUnicodeCharacter{2265}{\ensuremath{\geq}}
\usepackage{cuted}
\usepackage{url}

\def\BibTeX{{\rm B\kern-.05em{\sc i\kern-.025em b}\kern-.08em
    T\kern-.1667em\lower.7ex\hbox{E}\kern-.125emX}}

\begin{document}

\title{Evaluating the Reliability of Digital Forensic Evidence Discovered by Large Language Model: A Case Study \\

{\footnotesize }
\thanks{This material is based upon work supported in part by the National Science Foundation under Grant No. 2333949 and by the Bureau of Justice Assistance under Award No. 2019-DF-BX-K001.}
 }

\author{
\IEEEauthorblockN{Jeel Piyushkumar Khatiwala}
\IEEEauthorblockA{\textit{School of Criminal Justice} \\
\textit{College of Public Affairs}\\
\textit{University of Baltimore}\\
Maryland, USA \\
Jeel.khatiwala@ubalt.edu}
\and
\IEEEauthorblockN{Daniel Kwaku Ntiamoah Addai}
\IEEEauthorblockA{\textit{School of Criminal Justice} \\
\textit{College of Public Affairs} \\
\textit{University of Baltimore}\\
Maryland, USA\\
daniel.addai@ubalt.edu}
\and
\IEEEauthorblockN{Weifeng Xu}
\IEEEauthorblockA{\textit{School of Criminal Justice} \\
\textit{College of Public Affairs}\\
\textit{University of Baltimore}\\
Maryland, USA \\
wxu@ubalt.edu}
 }

\maketitle
 
\begin{abstract}  
The growing reliance on AI-identified digital evidence raises significant concerns about its reliability{,} particularly as large language models (LLMs) are increasingly integrated into forensic investigations.
This paper proposes a structured framework that automates forensic artifact extraction{,} refines data through LLM-driven analysis{,} and validates results using a Digital Forensic Knowledge Graph (DFKG).
Evaluated on a 13 GB forensic image dataset containing 61 applications{,} 2{,}864 databases{,} and 5{,}870 tables{,} the framework ensures artifact traceability and evidentiary consistency through deterministic Unique Identifiers (UIDs) and forensic cross-referencing.
We propose this methodology to address challenges in ensuring the credibility and forensic integrity of AI-identified evidence{,} reducing classification errors{,} and advancing scalable{,} auditable methodologies.
A comprehensive case study on this dataset demonstrates the framework’s effectiveness{,} achieving over 95\% accuracy in artifact extraction{,} strong support of chain-of-custody adherence{,} and robust contextual consistency in forensic relationships.
Key results validate the framework’s ability to enhance reliability{,} reduce errors{,} and establish a legally sound paradigm for AI-assisted digital forensics.
\end{abstract}


\section{Introduction}
The rapid expansion of digital data and the complexity of forensic investigations have created an urgent need for scalable, automated solutions to ensure the reliability, integrity, and legal admissibility of digital evidence \cite{Xu2019ForensicEvidence, Wu2020ForensicTools}. Large language models (LLMs) show strong potential in this area, offering capabilities for processing unstructured data, extracting entities, and performing contextual correlation \cite{Wickramasekara2024PotentialLLM, Scanlon2023ChatGPTForensics}. However, concerns persist regarding the authenticity, accountability, and forensic soundness of AI-generated outputs \cite{Xu2022PresentableEvidence}, stemming from biases in generated artifacts, inconsistent refinements, and the absence of structured validation.

A central challenge in existing forensic workflows is the lack of standardized frameworks for evidence discovery, artifact validation, and traceability \cite{Xu2022PresentableEvidence, Addai2024GraphEvidence}. Traditional methods often struggle with heterogeneous data such as call logs, messages, and encrypted databases, leading to incomplete extractions and insufficient context reconstruction. This impacts investigative outcomes, disrupts chain of custody, and raises questions of admissibility.

To address these issues, this paper presents a forensic evaluation framework that automates artifact discovery, extraction, and validation. It integrates advanced data extraction with LLM-driven refinement to improve accuracy and contextual clarity. A Digital Forensic Knowledge Graph (DFKG) models relationships among artifacts to support structured and verifiable analysis \cite{Karim2024GraphTheory}. The framework uses deterministic Unique Identifiers (UIDs) for artifact traceability, hypothesis testing against ground truth data, and forensic-specific metrics to assess evidentiary reliability.

This work addresses the following research questions:
\begin{itemize}
    \item \textbf{RQ1:} How can a standardized framework be implemented to evaluate the reliability of forensic evidence refined by LLMs using knowledge graph visualization?
    \item \textbf{RQ2:} How can the reliability of LLM-identified forensic evidence be measured in terms of accuracy, false positives, and artifact coverage?
    \item \textbf{RQ3:} What are the key sources of error or bias in LLM-generated forensic evidence, and how do they impact artifact classification?
    \item \textbf{RQ4:} How can LLM-assisted forensic evidence be validated to ensure traceability and accuracy through cross-referencing with original sources?
    \item \textbf{RQ5:} What best practices can enhance the admissibility and trustworthiness of LLM-derived evidence, focusing on traceability, standardized extraction, and validation workflows?
\end{itemize}

\noindent The contributions of this paper are as follows:
\begin{itemize}
    \item An automated forensic framework that unifies data extraction, LLM refinement, and graph-based correlation into a scalable pipeline.
    \item A transformation method that converts heterogeneous forensic databases into LLM-readable, audit-ready CSV format while preserving metadata for chain-of-custody compliance.
    \item A set of forensic-specific evaluation metrics, including Evidence Extraction Accuracy (EEA), Forensic Artifact Precision (FAP), Forensic Artifact Recall (FAR), Knowledge Graph Connectivity Accuracy (KGCA), Chain-of-Custody Adherence (CCA), and Contextual Consistency Score (CCS), for assessing AI-generated forensic outputs.
    \item A UID-based tagging system that assigns globally unique identifiers to each artifact, enabling traceability, validation, and auditability.
\end{itemize}

By addressing these challenges, this study advances digital forensic practice with a scalable, reliable, and legally defensible framework for AI-assisted investigations. The remainder of the paper is organized as follows: Section~\ref{sec:relatedwork} reviews related work; Section~\ref{sec:framework} outlines the proposed framework; Section~\ref{sec:casestudy} presents the case study and results; and Section~\ref{sec:conclusion} concludes with key insights and directions for future research.

\section{Related Work}
\label{sec:relatedwork}

The application of Artificial Intelligence (AI), particularly large language models (LLMs), in digital forensics has significantly advanced investigative workflows by automating artifact extraction, contextual analysis, and relationship visualization \cite{Wickramasekara2024PotentialLLM, Scanlon2023ChatGPTForensics}. These capabilities help address the increasing volume and complexity of digital evidence across platforms.

Traditional techniques such as regular expressions (regex) have long supported forensic preprocessing by identifying patterns in unstructured data. However, these approaches often require manual oversight and lack mechanisms to contextualize relationships, limiting their scalability for large-scale investigations \cite{Wu2020ForensicTools}. In contrast, LLMs can process semantically rich data with greater automation and contextual awareness.

Recent studies highlight LLMs’ potential in automating core forensic tasks, including entity lineage tracking and relationship identification. Wickramasekara et al. \cite{Wickramasekara2024PotentialLLM} demonstrated improved efficiency and traceability through LLM-based workflows, while specialized models such as ForensicLLM were introduced to increase domain-specific performance \cite{Xu2024AIForensics}.

Alongside LLMs, knowledge graphs (KGs) have emerged as powerful tools for modeling relationships between forensic artifacts. Xu and Xu \cite{Xu2022PresentableEvidence} proposed the Digital Forensic Knowledge Graph (DFKG) to support hypothesis-driven reasoning, and Addai et al. \cite{Addai2024GraphEvidence} explored multi-device relationship discovery. These graph-based frameworks help expose latent connections that traditional tools may overlook.

Graph theory continues to contribute to forensic linkage analysis at scale. Karim et al. \cite{Karim2024GraphTheory} demonstrated how graph algorithms can uncover critical relational structures. Pan et al. \cite{Pan2024LLM_KG_Roadmap} outlined integration strategies for combining LLMs with KGs to support hypothesis generation, and Wickramasekara et al. \cite{Wickramasekara2024HybridApproach} proposed hybrid models to enhance investigative outcomes in complex environments.

Additionally, Gen-AI platforms like ForenSift \cite{Talekar2024ForenSift} have introduced automation in digital forensics and incident response (DFIR). Built on the LangChain framework, ForenSift accelerates case analysis and improves artifact coverage through advanced entity correlation.

Despite these developments, challenges persist in scaling automated forensic analysis and maintaining contextual accuracy. Continued integration of LLMs with knowledge graphs, supported by structured forensic frameworks, offers a promising direction for enhancing the reliability and scalability of digital investigations \cite{Scanlon2023ChatGPTForensics, Wu2020ForensicTools}.

\begin{figure*}[htbp]
  \centering
  \includegraphics[width=475pt]{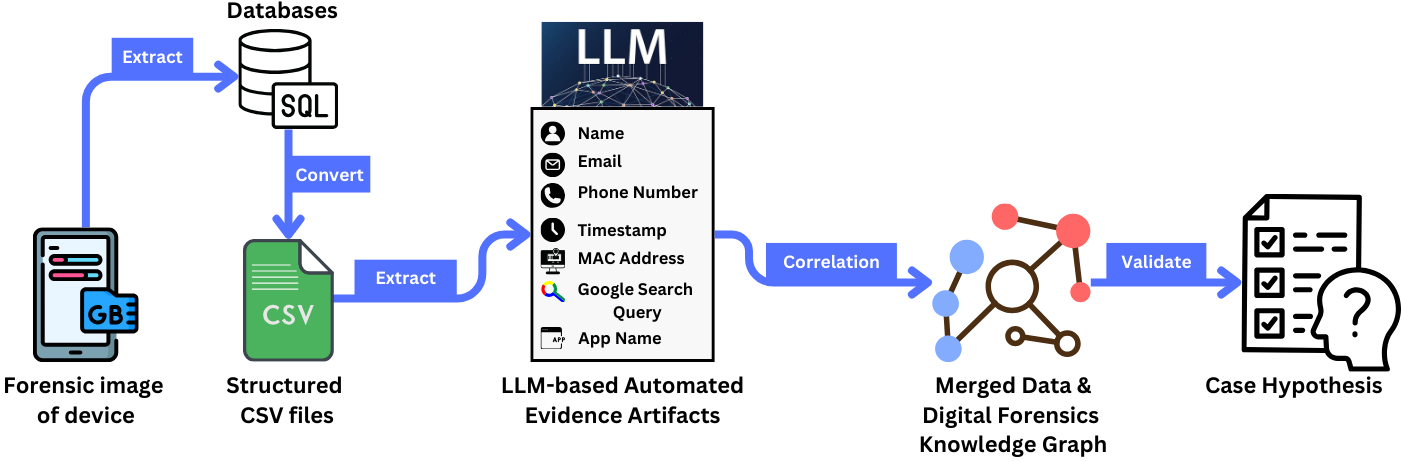}
  \caption{Proposed evaluation workflow: identifying artifacts containing digital evidence, transforming the digital evidence repository to an LLM-readable form, constructing the Digital Forensic Knowledge Graph (DFKG), and integrating LLM-refined evidence into the final graph representation.}
  \label{fig:workflow}
\end{figure*}

\section{Proposed Methodology}
\label{sec:framework}
We propose a structured framework to assess the reliability and accuracy of digital evidence extracted by large language models (LLMs). This framework ensures evidence integrity for both legal and investigative purposes, providing a holistic method for processing, analyzing, and validating forensic artifacts. As illustrated in Fig.~\ref{fig:workflow}, our methodology encompasses four key steps: (1) identifying forensic artifacts from mobile device databases (the ``digital evidence repository"), (2) transforming heterogeneous databases into a structured, LLM-readable CSV format, (3) constructing a Digital Forensic Knowledge Graph (DFKG) to model relationships among artifacts, and (4) applying robust forensic metrics to validate extracted evidence. By measuring forensic reliability at each stage, the framework enables end-to-end traceability and supports legally admissible digital forensic workflows \cite{Scanlon2023ChatGPTForensics, Wu2020ForensicTools}.

\subsection{Data Extraction and Preprocessing}
\subsubsection{Artifact Identification}
Artifact extraction focuses on mobile device databases, collectively termed the ``digital evidence repository,'' including call logs, messages, application records, and location data, essential for reconstructing user behavior. File signature analysis identifies and classifies databases based on unique byte patterns, enabling detection even when filenames or extensions are obfuscated. For example, databases such as \texttt{contacts2.db} (contacts), \texttt{sms.db} (messages), and \texttt{mmssms.db} (multimedia) are extracted from their respective directories. Each is exported as a CSV file with metadata (database name, table name, row number) preserved to ensure integrity and traceability \cite{Xu2022PresentableEvidence}.

\subsubsection{Data Transformation}
To support LLM-based processing, the evidence repository is converted into a flat CSV format. This transformation preserves the structural and semantic attributes of the original data, ensuring consistency and auditability, both essential for forensic authenticity. Each database table is exported as an individual CSV file containing metadata such as database name, table name, file path, and row number. Flattening relational data simplifies complex structures, resulting in a uniform format optimized for LLM analysis.

\subsection{Mathematical Formulations}
\subsubsection{Unique Identifier Generation (UID)}
A reliable Unique Identifier (UID) scheme ensures data integrity and verifiability, especially when identical databases exist across multiple locations. We generate a UID by applying a cryptographic hash to the \(\text{Device ID}\), \(\text{File Path}\), and \(\text{Database Name}\), then appending the \(\text{Table Name}\) and \(\text{Row Number}\). Truncating the hash to eight characters provides computational efficiency while maintaining uniqueness. The appended components ensure row-level granularity.

\begin{equation} 
\begin{aligned}
UID = \text{SHA-256}(\text{DeviceID} + \text{FilePath} + \text{DatabaseName})_{[:8]}\\
 + ``\_" + \text{Table} + ``\_" + \text{LID} 
\end{aligned}
\end{equation}
where:
\begin{itemize}
    \item \(\mathbf{Device ID}\): Represents the unique identifier of the source mobile device.
    \item  \(\mathbf{File Path}\): Indicates the exact location of the database file within the device.
    \item \(\mathbf{Database Name}\): Denotes the name of the extracted database file.
    \item \(\mathbf{Table}\): Identifies the table from which the forensic record is extracted.
    \item \(\mathbf{LID}\): Refers to the row number within the table to maintain intra-table uniqueness.
\end{itemize}

\paragraph{Example}
For instance, consider a WhatsApp database (\texttt{msgstore.db}) extracted from a device with:\vspace{0.5em}

\noindent\(\mathbf{(Device ID, \texttt{A1B2C3D4E5F6G7H8})}\) \\
\(\mathbf{(File Path, \texttt{/data/com.whatsapp/databases/})}\) \\
\(\mathbf{(Database Name, \texttt{msgstore.db})}\) \\
\(\mathbf{(T{name}, \texttt{messages})}\) \\
\(\mathbf{(LID, \texttt{42})}\) \\
\(\mathbf{(Hash Input, \texttt{A1B2C3D4E5F6G7H8 + /data/com.w}}\)\\
\(\mathbf{\texttt{hatsapp/databases/ + msgstore.db)}}\)\\
\(\mathbf{(Computed Hash, \texttt{788492af8249d22829c49c9...})}\) \\
\(\mathbf{(Extracted First 8 Characters, \texttt{788492af})}\) \\

\noindent The computed UID becomes:
\[
UID = \text{788492af\_messages\_42}
\]

This process ensures that identical databases from different sources generate distinct, verifiable UIDs.

\subsubsection{CSV Transformation for LLM Processing}
To ensure structured, traceable, and uniquely identifiable forensic data, each database table is converted into a CSV format, preserving metadata such as database name, table name, file path, UID, and row index alongside actual column values. Formally, a table row \( r' \) is defined as:
\begin{equation}
\begin{aligned}
CSV = \{ (\text{database}, \text{table}, \text{path}, \text{UID}, \text{LID}) \} \cup \{\text{(c, r'(c))}\\ \mid \text{c} \in \text{C}\}
\end{aligned}
\end{equation}
where:
\begin{itemize}
    \item  \(\mathbf{database}\) - Name of the database from which the table is extracted.
    \item  \(\mathbf{table}\) - Name of the table from which the row originates.
    \item  \(\mathbf{path}\) - The original file path of the database on the device.
    \item  \(\mathbf{UID}\) - Unique identifier ensuring global distinctiveness.
    \item \(\mathbf{LID}\) - Row index for intra-table differentiation.
    \item  \(\mathbf{(c, r'(c)) \mid c \in C}\) - Each column-value pair in the row, where \( C \) represents the set of column names.
\end{itemize}
Each row \(r'\) is a mapping \(r': A \to V\), where \(A\) is the set of attributes and \(V\) their corresponding values:
\begin{equation}
\begin{aligned}
r' = \{ (\text{database}, db), (\text{table}, t), (\text{path}, p), (\text{UID}, u), (\text{LID}, l) \}\\ \cup \{ \text{(c, r'(c))} \mid \text{c} \in \text{C}\}
\end{aligned}
\end{equation}

\paragraph{Example}
An entry from the \texttt{UserStore} table of \texttt{core.db} (Snapchat) might appear as: \vspace{0.5em}

\noindent\(\mathbf{(Database, \texttt{core.db})}\);\(\mathbf{(Table, \texttt{UserStore})}\);
\(\mathbf{(File Path, \texttt{...m.snapchat.android/databases/})}\);
\(\mathbf{(UID, \texttt{9f97eac5\_UserStore\_114})}\);
\(\mathbf{(LID, \texttt{114})}\);
\(\mathbf{(realval,\texttt{...x19heisenbergercarro@gmail.com-}}\)
\texttt{-\textbackslash x0...)}

All CSV data is merged into a unified row-based file, with each record represented as a set of attribute-value pairs. To manage data volume, the unified file is partitioned and every sixth row is sampled for reliability checks. Extraneous Android system and configuration metadata are excluded to prioritize user-centric forensic content, improving computational efficiency while maintaining integrity for LLM-based refinement and investigation.

\subsection{LLM-Assisted Artifact Refinement}
\label{sec:llm-refinement}

Once the repository is transformed into CSV format, large language models (LLMs) are applied to enhance the clarity and contextual accuracy of forensic artifacts. The LLM corrects inconsistencies, removes obfuscations, and reconstructs missing values. Each artifact remains linked to its source via its UID, enabling reliable consolidation across diverse data sources. Key entities, such as \textbf{Names}, \textbf{Phone Numbers}, \textbf{Emails}, \textbf{Timestamps}, \textbf{Application Names}, \textbf{Google Search Queries}, and \textbf{MAC Addresses}, are automatically extracted \cite{Wickramasekara2024PotentialLLM, Xu2024AIForensics}.

To standardize and validate these forensic artifacts, we employed the following LLM prompt:
\begin{strip}
\noindent
\begin{minipage}{\dimexpr\textwidth} 
\begin{lstlisting}[language=Python][
    numbersep=5pt,
    linenos,
    breaklines=true,
    frame=single,
    fontsize=\small,
    breaklines=true,
    bgcolor=white
]{text}
You are a forensic artifact refinement engine. Your task is to analyze each input row from a CSV file extracted from a mobile application database. Each row contains metadata and column-value pairs. Identify valid forensic artifacts, refine them, assign confidence scores, and output each artifact individually in the required format.
Each row contains:
Database Name (DB), Table Name (TN), File Path (FP), Row Line Number (LID), Unique Identifier (UID), One or more column-value pairs
Your task is to:
1) Use column names and metadata to determine the context of each value.
2) Extract only valid forensic artifacts of the following types:
     Email
     Phone Number
     Human Name (real human names)
     Username
     App Name (convert package names to recognizable names)
     Timestamp (convert to human-readable format)
     Search Keyword (from queries and titles)
     Message (user-generated text like SMS or chat)
     MAC Address
     Longitude
     Latitude
     Address (only identifiable physical locations)
3) For each artifact:
     Refine the value by correcting inconsistencies, removing obfuscations (e.g., encoded characters), and converting to a human-readable forensic format.
     Assign a confidence score between 1 (low certainty) and 10 (high certainty).
     Only retain artifacts with a confidence score of 5 or higher.
4) Output Structure (F):
    For every valid artifact identified, output a structured entry including:
     Entity Type (Exact label from the list below)
     Refined Value (The cleaned, normalized, human-readable artifact)
     Confidence Score (An integer from 1 to 10)
     Each extracted artifact must be listed separately, even if multiple artifacts are extracted from the same input row.
     In Output Use the exact entity type labels listed below for all output entries:
     App Name, Username, Human Name, Phone Number, Email, Search keyword, Message, MAC Address, Longitude, Latitude, Address, Timestamp.
5) Important rules:
     Do not include irrelevant system-level fields or Android internal configuration metadata
     Only use values that can be reliably interpreted based on column names and context
     Do not infer or invent artifact types not listed above
     If a value is partially recovered or decoded, include it only if confidence $\geq$ 5
     Each artifact must be written individually, not grouped or merged
        \end{lstlisting}
\end{minipage}
\end{strip}

To strengthen forensic reliability and limit speculative inferences, each entity is assigned a confidence score between 1 (least certain) and 10 (most certain). Artifacts with scores below a defined threshold (e.g., 5) are excluded during post-processing. This filtering mitigates the inclusion of low-certainty outputs, especially in cases involving incomplete metadata or fragmented records. Retained artifacts not only preserve chain-of-custody via UID, but also satisfy an added standard of inferential confidence, supporting evidentiary integrity.

\paragraph{Example}
Consider an entry from the \texttt{UserStore} table of \texttt{core.db} (Snapchat). After CSV conversion, it may appear as: \vspace{0.5em}

\noindent\(\mathbf{(Database, \texttt{core.db})}\);\(\mathbf{(Table, \texttt{UserStore})}\);
\(\mathbf{(File Path, \texttt{...m.snapchat.android/databases/})}\);
\(\mathbf{(UID, \texttt{9f97eac5\_UserStore\_114})}\);
\(\mathbf{(LID, \texttt{114})}\);
\(\mathbf{(realval,\texttt{...x19heisenbergercarro@gmail.com-}}\)
\texttt{-\textbackslash x0...)}

After processing by the LLM using the above prompt, the refined artifacts are stored in separate CSV files:
\begin{itemize}[leftmargin=*, label=\textbullet]
\item \textbf{Emails.csv}: \texttt{heisenbergcarro@gmail.com}, Confidence: \texttt{10}
\item \textbf{Name.csv}: \texttt{Heisenberg White}, Confidence: \texttt{7}
\item \textbf{AppNames.csv}: \texttt{Snapchat}, Confidence: \texttt{9}
\end{itemize}

Each refined artifact retains the original UID, maintaining chain-of-custody. If an artifact, such as an obfuscated search query or partially reconstructed name, receives a confidence score below 5, it is excluded to ensure only high-integrity evidence is used for further forensic correlation.

\subsection{Knowledge Graph Construction}

Following LLM-based refinement, artifacts from separate CSV files are aligned using UIDs. Entries sharing a UID are merged into consolidated records, integrating complementary attributes (e.g., names, timestamps, application references). This reduces duplication while preserving contextual integrity required for the construction of the Digital Forensic Knowledge Graph (DFKG).

The DFKG employs a node-edge model, where nodes represent forensic entities (e.g., email address, timestamp) and edges define relationships (e.g., communication links, temporal correlations). The summary nodes, shown in Table~\ref{Summary_Node}, trace the transformation from raw input to refined output, strengthening traceability and auditability.

Isolated nodes, those without direct links, are grouped alphabetically by application name. Since all applications originate from the same device, this enhances readability and supports intra-device analysis without implying semantic linkage. It aids contextualization while preserving forensic independence.

To ensure data quality, only artifacts with a score of confidence ≥5 are retained in the DFKG. This threshold, set during LLM refinement, filters out low-certainty outputs from fragmented metadata. As a result, the graph emphasizes high-confidence relationships, minimizing false associations and supporting hypothesis validation.

Each node and edge is tied to a single device via deterministic UIDs, ensuring cross-device separation and preserving evidentiary integrity in accordance with chain-of-custody principles.

\begin{figure*}[!hbt]
    \centering
    \fbox{%
        \includegraphics[width=475pt]{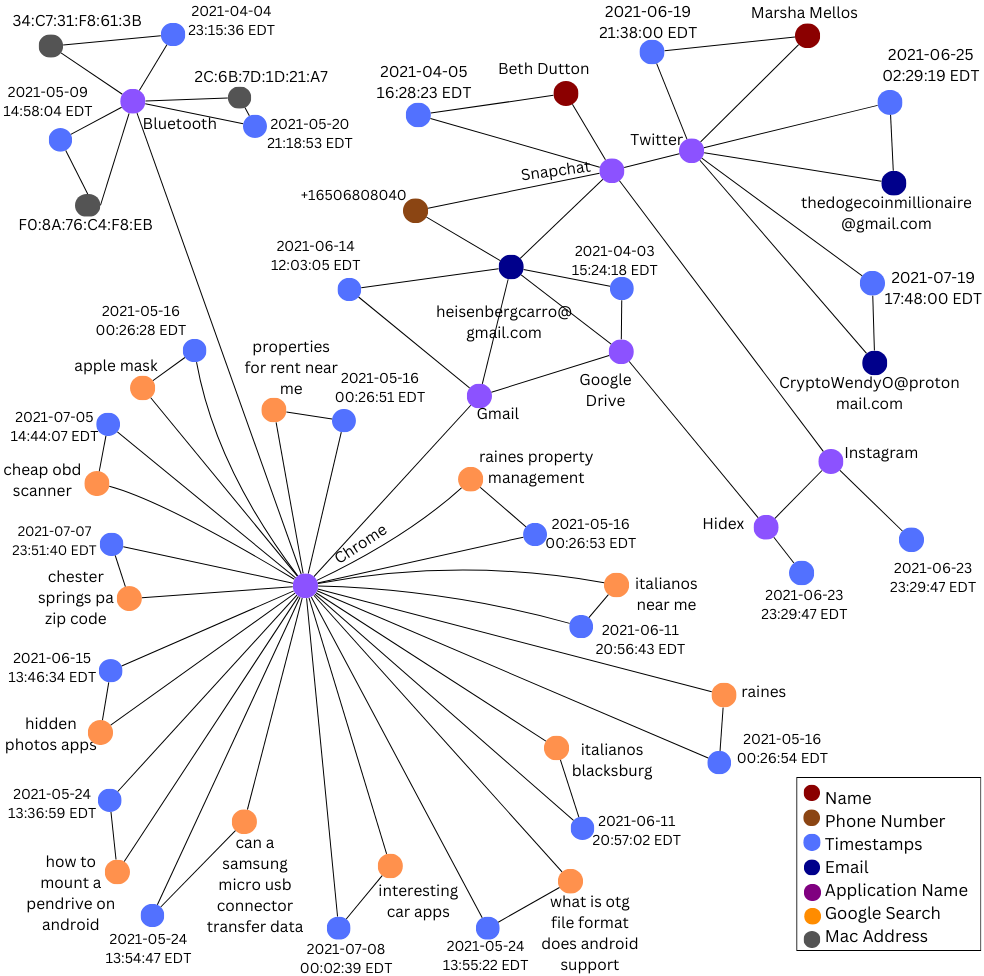}
    }
    \caption{Visualization of the DFKG with Artifact Relationships}
    \label{fig:graph}
\end{figure*}

An example graph is shown in Fig.\ref{fig:graph}. The DFKG provides a unified structure for modeling entities and relationships, supporting scalable, auditable forensic investigations. Its UID-based alignment and contextual linkage enhance clarity and uphold forensic standards\cite{Xu2022PresentableEvidence, Xu2024AIForensics}.

While structurally similar to threat intelligence formats such as STIX~\cite{OASIS2020STIX}, the DFKG is independently developed. Its design prioritizes reproducibility, traceability, and forensic provenance to support legal validation. Future work will explore exporting UID-linked entities in STIX-compatible formats for interoperability across forensic and intelligence platforms.

\subsection{Evaluation Metrics}
\label{sec:evaluation}

We evaluated the proposed forensic framework using ground truth data from Cellebrite's Capture-the-Flag (CTF) competitions \cite{Cellebrite2021CTF, Cellebrite2022CTF}, validating results against Cellebrite solutions and expert forensic analysis. A total of 2,000 rows were processed. The LLM-based extraction initially identified 42 potential evidence items, of which 40 were correct and 2 were false positives. After UID-based consolidation, these were merged into 26 distinct evidence records, with 24 correct and 2 incorrect records. The final Digital Forensic Knowledge Graph (DFKG) constructed from these records contained 72 connections, of which 68 accurately represented true relationships.

\begin{table*}[!htbp]
\renewcommand{\arraystretch}{1.2}
\caption{Summary of the End-to-End Workflow for Extracting and Validating Forensic Artifacts}
\begin{center}
\begin{tabular}{|p{2.5cm}|p{6cm}|p{3.5cm}|p{2.5cm}|}
\hline
\textbf{\centerline{Artifact Type}} & \textbf{\centerline{Database Source (Column, Row Data)}} & \textbf{\centerline{Extracted Artifacts}} & \textbf{\centerline{Destination CSV file}} \\ \hline

Email & (realVal,...a\textbackslash x19heisenbergercarro@gmail.com\textbackslash x0...) & heisenbergercarro@gmail.com & Email.csv \\ \hline
Timestamp & (proto,…x03\textbackslash x0f``1617477858090"\textbackslash xe2\textbackslash x03\textbackslash x…) & 03 April 2021 15:24:18 & Timestamp.csv \\ \hline
Phone Number & (blobVal,…n\textbackslash x0c+16506808040\textbackslash x12\textbackslash t\textbackslash n\textbackslash x01…) & +16506808040 & Phone\_number.csv \\ \hline
Name & (Data,…\texttt"ull\_name\texttt":\texttt"Marsha Mellos\texttt",\texttt"pro…) & Marsha Mellos & Name.csv \\ \hline
Google Search Query & (title, hidden photos apps - Google Search) & hidden photos apps & Google\_Search.csv \\ \hline
MAC Address & (address, 34:C7:31:F8:61:3B) & 34:C7:31:F8:61:3B & Mac\_addr.csv \\ \hline
App Name & (DPath, …ser/0/com.instagram.android/databases/…) & Instagram & Appname.csv \\ \hline
\end{tabular}
\label{Summary_Node}
\vspace{2pt}
\end{center}
{\footnotesize 
\textbf{Note:} 
This table summarizes the end-to-end flow for data from Database Source to Extracted Artifacts.\\
\textbf{Abbreviations:}  
\textit{Artifact Type} = Type of forensic artifact; \textit{Process} = Technique used for artifact extraction.
}
\end{table*}

\vspace{1em}
\textbf{Metric 1: Evidence Extraction Accuracy (EEA)}  
\[
\text{EEA} = \frac{\text{True Extractions}}{\text{Total Potential Extractions}} \times 100\%
\]
\[
\text{EEA} = \frac{40}{42} \times 100\% \approx 95.24\%
\]
\textbf{Explanation:} EEA quantifies the accuracy of the initial LLM-based extraction process.

\vspace{1em}
\textbf{Metric 2: Evidence Consolidation Accuracy (ECA)}  
\[
\text{ECA} = \frac{\text{Correctly Consolidated Records}}{\text{Total Consolidated Records}} \times 100\%
\]
\[
\text{ECA} = \frac{24}{26} \times 100\% \approx 92.31\%
\]
\textbf{Explanation:} ECA reflects the effectiveness of merging overlapping evidence items using UID.

\vspace{1em}
\textbf{Metric 3: Knowledge Graph Connectivity Accuracy (KGCA)}  
\[
\text{KGCA} = \frac{\text{Correctly Established Connections}}{\text{Total Connections}} \times 100\% 
\]
\[
\text{KGCA} = \frac{68}{72} \times 100\% \approx 94.44\%
\]
\textbf{Explanation:} KGCA measures the accuracy of relationships formed in the final knowledge graph.

\vspace{1em}
\textbf{Metric 4: Forensic Artifact Precision (FAP)}  
\[
\text{FAP} = \frac{\text{TP}}{\text{TP} + \text{FP}} \times 100\% = \frac{40}{40 + 2} \times 100\% \approx 95.24\%
\]
\textbf{Explanation:} FAP quantifies the proportion of correctly identified artifacts among all identified artifacts.

\vspace{1em}
\textbf{Metric 5: Forensic Artifact Recall (FAR)}  
Assuming no relevant evidence was missed (FN = 0):
\[
\text{FAR} = \frac{\text{TP}}{\text{TP} + \text{FN}} \times 100\% = \frac{40}{40 + 0} \times 100\% = 100\%
\]
\textbf{Explanation:} FAR measures the system's ability to extract all relevant evidence items.

\vspace{1em}
\textbf{Metric 6: Artifact Integrity Score (AIS)}  
\[
\text{AIS} = \frac{\text{Correctly Extracted Artifacts}}{\text{Total Extracted Artifacts}} \times 100\%
\]
\[
\text{AIS} = \frac{40}{42} \times 100\% \approx 95.24\%
\]
\textbf{Explanation:} AIS evaluates fidelity to original evidence based on ground truth.

\vspace{1em}
\textbf{Metric 7: Chain of Custody Adherence (CCA)}  
\[
\text{CCA} = 100\%
\]
\textbf{Explanation:} CCA confirms that all artifacts retain key metadata (e.g., UID), preserving legal admissibility.

\vspace{1em}
\textbf{Metric 8: Contextual Consistency Score (CCS)}  
\[
\text{CCS} = \frac{\text{Artifacts Matching Expected Contexts}}{\text{Total Extracted Artifacts}} \times 100\%
\]
\[
\text{CCS} = \frac{40}{40} \times 100\% = 100\%
\]
\textbf{Explanation:} CCS measures how well extracted evidence aligns with expected forensic context.

\noindent
The initial LLM-based extraction achieved an EEA of 95.24\%, identifying 42 artifacts with 40 correct and 2 false positives. UID-based consolidation merged these into 26 records with an ECA of 92.31\%. The resulting DFKG formed 72 connections, of which 68 were accurate, yielding a KGCA of 94.44\%. Although these metrics reflect the complete LLM output, a post-processing confidence threshold of 5 (on a 1 to 10 scale) was applied to exclude low-certainty artifacts from downstream analysis and graph construction. This safeguard enhances reliability by filtering speculative inferences and reinforcing evidentiary integrity.

\section{Case Study: Mobile Phone Forensics}
\label{sec:casestudy}

This section presents a case study applying the proposed forensic framework to assess the reliability and soundness of extracted digital evidence in a mobile phone investigation. The evaluation is structured around five research questions, each addressing a distinct aspect of the forensic pipeline, including artifact identification, data transformation, LLM refinement, knowledge graph construction, and overall integrity.

\subsection{Case Study Description}
The case centers on a suspected auto theft and resale scheme involving three individuals: Heisenberg, Beth Dutton, and Marsha Mellos. On July 21\textsuperscript{st}, law enforcement apprehended Beth at a restaurant in Vienna, VA, following an invitation from Heisenberg. During questioning, Beth asserted that both she and her sister Marsha were innocent, claiming Heisenberg was the orchestrator behind the operation. The sisters, who operate a cattle business in Montana, stated they were unintentionally linked to his activities.

The primary source of digital evidence was Heisenberg’s Android smartphone, seized to evaluate his potential involvement. Key indicators included Google searches related to auto theft, Bluetooth connections to vehicles, and communications involving all three individuals, each contributing context regarding Heisenberg’s actions and associations.

To test the framework, a forensic image of a Samsung Galaxy Note 10 from Cellebrite’s 2021 and 2022 Capture-the-Flag (CTF) competitions was used \cite{Cellebrite2021CTF, Cellebrite2022CTF}. The 13 GB image included data from 61 applications (both user-installed and system), 2,864 databases, and 5,870 tables. The framework was applied to automate artifact extraction, normalize records, refine values using OpenAI’s GPT-4, and conduct semantic analysis with emphasis on metadata integrity, accountability, and traceability.

Entity relationships were modeled using the Digital Forensic Knowledge Graph (DFKG), facilitating systematic event reconstruction and uncovering connections between the suspects. This case study demonstrates the framework’s capability to automate digital evidence discovery, improve artifact accountability, and support dependable forensic investigations. Performance metrics, including accuracy, precision, recall, F1-score, and legal admissibility (Section~\ref{sec:evaluation}), further validate the reliability of LLM-assisted evidence processing under realistic conditions.

\begin{table*}[!htbp]
\renewcommand{\arraystretch}{1.2} 
\caption{Explanation and Validation of Hypotheses for Edge Types in the DFKG}
\begin{center}
\begin{tabular}{|c|p{12cm}|c|c|c|}
\hline
\textbf{Edge Type} & \textbf{\centerline{ Example of Hypothesis}} & \textbf{Total} & \textbf{Valid} & \textbf{Invalid} \\ \hline

\multirow{3}{*}{\begin{tabular}[c]{@{}c@{}}Timestamps and\\ Application Name\end{tabular}} 
& User logged into Google Drive on 2021-04-03 at 15:24:18 EDT. & \multirow{3}{*}{24} & \multirow{3}{*}{22} & \multirow{3}{*}{2} \\ \cline{2-2}
& User searched for “interesting car apps” in Chrome on 2021-07-08 00:02:39 EDT. & & & \\ \cline{2-2}
& User initiated a Bluetooth connection on 2021-05-20 at 21:18:53 EDT. & & & \\ \hline

\multirow{3}{*}{\begin{tabular}[c]{@{}c@{}}Email and\\ Application Name\end{tabular}} 
& User accessed Google Drive associated with the email heisenbergcarro@gmail.com. & \multirow{3}{*}{5} & \multirow{3}{*}{3} & \multirow{3}{*}{2} \\ \cline{2-2}
& User account associated with the email heisenbergcarro@gmail.com was linked to Snapchat. & & & \\ \cline{2-2}
& The Twitter account associated with the email CryptoWendyO@protonmail.com is linked to Heisenberg’s Twitter interactions. & & & \\ \hline

\multirow{3}{*}{\begin{tabular}[c]{@{}c@{}}Application Name\\ and Google\\ Search Query\end{tabular}} 
& User performed a Google search for ``italianos near me" in Chrome. & \multirow{3}{*}{13} & \multirow{3}{*}{13} & \multirow{3}{*}{0} \\ \cline{2-2}
& User performed a Google search for ``hidden photos apps" in Chrome. & & & \\ \cline{2-2}
& User performed a Google search for ``interesting car apps" in Chrome. & & & \\ \hline

\multirow{3}{*}{\begin{tabular}[c]{@{}c@{}}MAC Address\\ and Application\\ Name\end{tabular}} 
& Device with MAC address 34:C7:31:F8:61:3B was connected via Bluetooth. & \multirow{3}{*}{3} & \multirow{3}{*}{3} & \multirow{3}{*}{0} \\ \cline{2-2}
& Device with MAC address 2C:6B:7D:1D:21 was connected via Bluetooth. & & & \\ \cline{2-2}
& Device with MAC address F0:8A:76:C4:F8 was connected via Bluetooth. & & & \\ \hline

\multirow{2}{*}{\begin{tabular}[c]{@{}c@{}}Timestamps and\\ Email\end{tabular}} 
& User interacted with CryptoWendyO@protonmail.com content on Twitter at 2021-07-19 17:48 EDT. & \multirow{2}{*}{5} & \multirow{2}{*}{5} & \multirow{2}{*}{0} \\ \cline{2-2}
& User interacted with content associated with the email thedogecoinmillionaire@gmail.com on Twitter on 2021-06-25 02:29:19 EDT. & & & \\ \hline

\multirow{3}{*}{\begin{tabular}[c]{@{}c@{}}Timestamp and\\ Google Search\\ Query\end{tabular}} 
& User searched for ``chester springs pa zip code" at 2021-07-07 23:51:40 EDT. & \multirow{3}{*}{13} & \multirow{3}{*}{13} & \multirow{3}{*}{0} \\ \cline{2-2}
& User searched for ``hidden photos apps" at 2021-06-15 13:46:34 EDT. & & & \\ \cline{2-2}
& User searched for ``interesting car apps" at 2021-07-08 00:02:39 EDT. & & & \\ \hline

\multirow{3}{*}{\begin{tabular}[c]{@{}c@{}}Timestamp and\\ MAC Address\end{tabular}} 
& Device with MAC 34:C7:31:F8:61:3B connected via Bluetooth on 2021-04-04 at 23:15:36 EDT. & \multirow{3}{*}{3} & \multirow{3}{*}{3} & \multirow{3}{*}{0} \\ \cline{2-2}
& Device with MAC F0:8A:76:C4:F8 connected via Bluetooth on 2021-05-09 at 14:58:04 EDT. & & & \\ \cline{2-2}
& Device with MAC 2C:6B:7D:1D:21 connected via Bluetooth on 2021-05-20 at 21:18:53 EDT. & & & \\ \hline

\multirow{2}{*}{\begin{tabular}[c]{@{}c@{}}Name and\\ Timestamp\end{tabular}} 
& User interacted with Beth Dutton on Snapchat on 2021-04-05 at 16:28:23 EDT. & \multirow{2}{*}{2} & \multirow{2}{*}{2} & \multirow{2}{*}{0} \\ \cline{2-2}
& User interacted with Marsha Mellos on Twitter on 2021-06-19 at 21:38:00 EDT. & & & \\ \hline

\multirow{2}{*}{\begin{tabular}[c]{@{}c@{}}Name and\\ Application Name\end{tabular}} 
& User interacted with Beth Dutton on Snapchat. & \multirow{2}{*}{2} & \multirow{2}{*}{2} & \multirow{2}{*}{0} \\ \cline{2-2}
& User interacted with Marsha Mellos on Twitter. & & & \\ \hline

\begin{tabular}[c]{@{}c@{}}Phone Number\\ and Application\\ Name\end{tabular} 
& The phone number +16506808040 is associated with Snapchat. & 1 & 1 & 0 \\ \hline

\begin{tabular}[c]{@{}c@{}}Phone Number\\ and Email\end{tabular} 
& The email heisenbergcarro@gmail.com is associated with the phone number +16506808040 on Snapchat. & 1 & 1 & 0 \\ \hline

\textbf{Total} & & \textbf{72} & \textbf{68} & \textbf{4} \\ \hline
\end{tabular}
\label{hypgothesis}
\vspace{2pt}
\end{center}
{\footnotesize 
\textbf{Note:} 
\textit{This table provides insights into the validation of hypotheses for relationships between artifact types in the Digital Forensic Knowledge Graph (DFKG).}\\
\textbf{Abbreviations:} 
\textit{Total} = total number of hypotheses tested; \textit{Valid} = number of validated hypotheses; \textit{Invalid} = number of hypotheses that failed validation.
}
\end{table*}

\subsection{RQ 1: How can a standardized framework be implemented to evaluate the reliability of forensic evidence refined by LLM using visualization of knowledge graphs?}

The proposed methodology introduces a structured framework for evaluating the reliability, accuracy, and forensic validity of digital evidence refined by large language models (LLMs). This is achieved by integrating database extraction, structured transformation, artifact refinement, and Digital Forensic Knowledge Graph (DFKG) visualization.

SQLite databases, commonly used to store communication logs, app usage records, and geolocation data, are extracted from forensic images \cite{Scanlon2023ChatGPTForensics}. These are converted into structured CSV files while preserving metadata such as database name, table name, and row index. A deterministic Unique Identifier (UID) scheme (see Section~\ref{sec:framework}) ensures traceability and integrity.

Each row in the CSV is then processed by an LLM module, which extracts and refines key entities, including \texttt{Name}, \texttt{Phone Number}, \texttt{Email}, \texttt{Timestamps}, and \texttt{Application Name}. The model corrects inconsistencies and obfuscations while maintaining UID-based linkage to the original data.

Refined artifacts are embedded into the DFKG, where nodes represent entities and edges capture contextual relationships. This graph-based structure supports evidence correlation, hypothesis testing, and contextual validation. Ground-truth datasets from Cellebrite’s CTF competitions \cite{Cellebrite2021CTF, Cellebrite2022CTF} are used for evaluation, with records categorized as true positives, false positives, or false negatives.

To assess performance, the framework employs a suite of forensic-specific metrics: Evidence Extraction Accuracy (EEA), Forensic Artifact Precision (FAP), Forensic Artifact Recall (FAR), Knowledge Graph Connectivity Accuracy (KGCA), Chain of Custody Adherence (CCA), and Contextual Consistency Score (CCS) (Section~\ref{sec:evaluation}). These metrics validate both artifact authenticity and contextual traceability.

By Combining UID traceability, LLM refinement, graph structuring, and hypothesis validation, the framework enables transparent and reproducible evaluation of forensic evidence in AI-assisted investigations.

\subsection{RQ 2: How can the reliability of LLM-identified forensic evidence be measured in terms of accuracy, false positives, and artifact coverage?}

The reliability of LLM-assisted forensic evidence is assessed using forensic-specific metrics, including accuracy, false positive rates, and artifact coverage. These metrics evaluate the framework’s ability to extract, validate, and refine forensic artifacts while minimizing classification errors. The empirical results confirm the framework’s effectiveness in ensuring forensic integrity.

\textbf{Evidence Extraction Accuracy (EEA)} measures the proportion of correctly identified forensic artifacts relative to the total extractions. Out of \textbf{42 potential evidence items}, \textbf{40 were correctly identified}, with \textbf{2 false positives}, resulting in an EEA of 95.24\%, demonstrating the framework’s reliability in initial LLM-based extraction.

\textbf{Forensic Artifact Precision (FAP)} quantifies the correctness of positively identified artifacts. Since \textbf{40 out of 42 extracted artifacts were correctly classified}, the precision score is 95.24\%, confirming that the majority of classified artifacts were forensically valid.

\textbf{Forensic Artifact Recall (FAR)} evaluates the framework’s ability to detect all relevant forensic artifacts. Since no relevant evidence was missed (false negatives = 0), the recall was measured at 100\%, ensuring complete forensic data retrieval.

\textbf{Forensic Artifact F1-Score (FAF1)} balances precision and recall, calculated as 97.56\%, reinforcing the system’s reliability in forensic artifact classification. Specifically, the LLM-based framework achieved a precision of 95.24\%, a recall of 100\%, and an F1-score of 97.56\%, highlighting its robust performance across key forensic metrics.

\textbf{Evidence Consolidation Accuracy (ECA)} assesses the merging of overlapping forensic evidence based on Unique Identifiers (UIDs). Initially, \textbf{42 evidence items} were identified, but after UID-based consolidation, they were merged into 26 distinct forensic records, with 24 correctly consolidated and 2 incorrectly merged. This results in an ECA of 92.31\%, highlighting the effectiveness of UID-driven consolidation.

\textbf{Knowledge Graph Connectivity Accuracy (KGCA)} measures the integrity of relationships within the Digital Forensic Knowledge Graph (DFKG). The forensic graph initially contained 72 established connections, out of which 68 were correctly mapped, leading to a KGCA of 94.44

\textbf{Contextual Consistency Score (CCS)} validates the alignment of extracted forensic artifacts with expected forensic patterns. The system achieved 100\% CCS, ensuring that all refined forensic evidence maintained consistency with expected investigative scenarios.

\textbf{Chain-of-Custody Adherence (CCA)} ensures the preservation of metadata integrity throughout the forensic process. The system maintained 100\% CCA, confirming that all forensic artifacts retained their original metadata, ensuring traceability and legal admissibility.

These findings confirm that the proposed forensic framework achieves high artifact extraction accuracy, precise consolidation, strong forensic graph relationship integrity, and full traceability. The results validate its ability to maintain forensic authenticity and enhance investigative reliability.

\subsection{RQ 3: What are the key sources of error or bias in LLM-generated forensic evidence and how do they impact artifact classification?}

The framework identifies key sources of error and bias that affect the reliability of LLM-generated forensic evidence. These include incomplete metadata, fragmented or deleted records, and contextual inference bias. Each impacts how digital evidence is interpreted, particularly when relationships are inferred by large language models (LLMs).

\begin{figure}[htbp]
\centerline{\fbox{\includegraphics[scale=0.35]{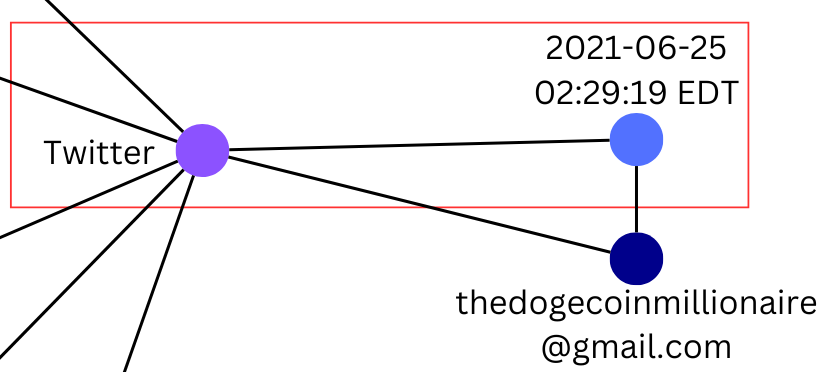}}}
\caption{Invalid hypothesis for Timestamps and Application Names.}
\label{fig:ICT1}
\end{figure}

\textbf{Incomplete metadata} weakens classification by omitting essential context like user IDs, session logs, or app usage traces. Artifacts may appear syntactically correct, yet lack the contextual fields needed for accurate association in the Digital Forensic Knowledge Graph (DFKG). As shown in Figure~\ref{fig:ICT1}, the timestamp \texttt{2021-06-25 02:29:19 EDT} was mistakenly linked to \texttt{Twitter} due to missing metadata. This misclassification was only identified during graph-based validation against Cellebrite’s ground truth, despite an Evidence Extraction Accuracy (EEA) of 95.24\% (40 of 42 correct).

\textbf{Fragmentation and deleted data} disrupt artifact reconstruction by stripping structural cues. For example, a corrupted email \texttt{19xxheisenbergcarro@gmail.comx1} was initially misclassified. Although the LLM accurately refined it to \texttt{heisenbergcarro@gmail.com}, insufficient metadata prevented reliable linkage to an application. This relationship was flagged during DFKG validation and contributed to two false positives in Forensic Artifact Precision (FAP), recorded at 95.24\%.

\begin{figure}[htbp]
\centerline{\fbox{\includegraphics[scale=0.35]{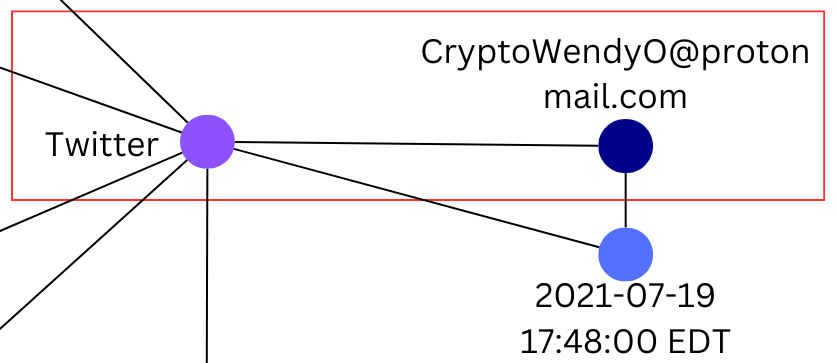}}}
\caption{Invalid hypothesis for Email and Application Names.}
\label{fig:ICT2}
\end{figure}

\textbf{Contextual bias in LLM inference} occurs when relationships are inferred from co-occurrence rather than substantiated connections. While malformed entries like \texttt{marketing@get@upside} were correctly filtered, subtler errors persisted. Figure~\ref{fig:ICT2} shows one such case where \texttt{CryptoWendyO@protonmail.com} was wrongly linked to \texttt{Twitter}, despite the absence of corroborating logs. This contributed to a Knowledge Graph Connectivity Accuracy (KGCA) of 94.44\% (68 of 72 valid relationships).

These false positives emerged only during graph validation. To mitigate such issues, the framework assigns confidence scores (1–10) during refinement and excludes artifacts below a threshold of 5. This removes speculative outputs before graph construction. Manual graph-based hypothesis testing serves as a secondary safeguard, revealing context-related errors that automated filters may miss. Although all metrics include pre-filtering data for transparency, this dual-layer validation improves overall forensic reliability.

\subsection{RQ 4: How can LLM-assisted forensic evidence be validated to ensure traceability and accuracy through cross-referencing with sources?}

The validation of LLM-assisted forensic evidence is achieved through deterministic Unique Identifiers (UIDs) and structured cross-referencing within the Digital Forensic Knowledge Graph (DFKG). Each artifact is assigned a UID derived from key attributes, including the database name, table name, file path, and row index, preserving forensic provenance and ensuring end-to-end auditability. This approach safeguards the chain of custody, enabling investigators to trace refined artifacts back to their original raw sources.

Empirical validation confirms that every artifact in the DFKG retains its UID throughout the forensic pipeline, allowing discrepancies introduced during LLM processing to be detected. The framework achieved an Evidence Consolidation Accuracy (ECA) of 92.31\%, correctly merging 24 out of 26 consolidated forensic records, ensuring UID-based integrity. Additionally, it maintained 100\% Chain-of-Custody Adherence (CCA), verifying that all extracted artifacts preserved their original metadata. The Knowledge Graph Connectivity Accuracy (KGCA) of 94.44\% further validated the integrity of forensic relationships.

While two artifacts were incorrectly consolidated, UID-driven cross-referencing identified these inconsistencies, demonstrating the system’s ability to correct forensic errors. By embedding UID-based traceability, enforcing structured cross-referencing, and ensuring consistent metadata preservation, the framework maintains evidentiary integrity, supporting a legally admissible and auditable forensic methodology.

\subsection{RQ 5: What best practices can improve the admissibility and trustworthiness of LLM-derived evidence, focusing on traceability, standardized extraction, and validation workflows?}

A structured forensic workflow enhances the admissibility and trustworthiness of LLM-derived evidence. In this study, a 13~GB forensic image containing 61 applications, 2,864 databases, and 5,870 tables was converted into structured CSV files, preserving comprehensive metadata and enabling artifact traceability.

Provenance is maintained using deterministic Unique Identifiers (UIDs), generated by SHA-256 hashing of device ID, database name, file path, table name, and row index. This achieves 100\% Chain-of-Custody Adherence (CCA), ensuring each artifact links to its original source.

Standardized CSV exports ensure consistent formatting, enabling LLMs to reconstruct fragmented artifacts. For instance, the obfuscated email address \texttt{19xxheisenbergcarro@gmail.comx1} was accurately refined to \texttt{heisenbergcarro@gmail.com}, contributing to a Forensic Artifact Precision (FAP) of 95.24

Validation confirmed an Evidence Extraction Accuracy (EEA) of 95.24\% (40 out of 42 correct) and a Knowledge Graph Connectivity Accuracy (KGCA) of 94.44\%, affirming accurate relationship mapping. The framework adheres to ISO/IEC 27037 guidelines on reproducibility and authenticity, supporting its legal admissibility.

In summary, practices such as standardized extraction, UID-based traceability, LLM refinement, and robust cross-referencing ensure that LLM-derived forensic evidence is reliable, reproducible, and legally defensible.

\section{Conclusion and Future Work}
\label{sec:conclusion}

This study presented a structured framework to evaluate the reliability and forensic soundness of digital evidence refined by large language models (LLMs). Through hypothesis validation, UID-based cross-referencing, and Digital Forensic Knowledge Graph (DFKG) construction, the framework ensures artifact integrity, traceability, and contextual accuracy in AI-assisted investigations.

Empirical results demonstrated strong forensic performance: 95.24\% Evidence Extraction Accuracy (EEA), 95.24\% Forensic Artifact Precision (FAP), and 100\% Forensic Artifact Recall (FAR), indicating minimal false positives and complete artifact retrieval. The system also achieved a 97.56\% Forensic Artifact F1-Score (FAF1), 94.44\% Knowledge Graph Connectivity Accuracy (KGCA), and 100\% scores for both Chain-of-Custody Adherence (CCA) and Contextual Consistency Score (CCS), supporting legal admissibility and reliability.

Key challenges include handling ambiguous artifacts from incomplete logs, fragmented metadata, or static databases. These conditions may affect relationship reconstruction and require enhanced validation and broader cross-referencing.

The current confidence-based filtering mechanism, though effective in minimizing false positives, may omit valid artifacts lacking sufficient context due to deletion, encryption, or corruption. Future work will explore adaptive thresholding, contextual re-validation, and deferred artifact review to mitigate premature exclusions.

Ongoing development aims to scale the framework for multi-device investigations and assess its performance in real-world cybercrime scenarios. Planned integration with AI-driven forensic reasoning systems will further improve classification and timeline reconstruction.

To support interoperability, future versions will export UID-linked artifacts using Structured Threat Information Expression (STIX)~\cite{OASIS2020STIX}. Although the current design prioritizes deterministic identifiers for auditability, STIX-compliant output will enable standardized evidence exchange across forensic tools and collaborative environments.



\bibliographystyle{IEEEtran}  
\bibliography{bibliography/references} 

\begin{thebibliography}{10}
\providecommand{\url}[1]{#1}
\csname url@samestyle\endcsname
\providecommand{\newblock}{\relax}
\providecommand{\bibinfo}[2]{#2}
\providecommand{\BIBentrySTDinterwordspacing}{\spaceskip=0pt\relax}
\providecommand{\BIBentryALTinterwordstretchfactor}{4}
\providecommand{\BIBentryALTinterwordspacing}{\spaceskip=\fontdimen2\font plus
\BIBentryALTinterwordstretchfactor\fontdimen3\font minus \fontdimen4\font\relax}
\providecommand{\BIBforeignlanguage}[2]{{%
\expandafter\ifx\csname l@#1\endcsname\relax
\typeout{** WARNING: IEEEtran.bst: No hyphenation pattern has been}%
\typeout{** loaded for the language `#1'. Using the pattern for}%
\typeout{** the default language instead.}%
\else
\language=\csname l@#1\endcsname
\fi
#2}}
\providecommand{\BIBdecl}{\relax}
\BIBdecl

\bibitem{Xu2019ForensicEvidence}
\BIBentryALTinterwordspacing
W.~Xu, J.~Yan, and H.~Chi, ``A forensic evidence acquisition model for data leakage attacks,'' in \emph{Proc. 17th IEEE Int. Conf. Intelligence and Security Informatics (ISI)}, 2019, pp. 53--58. [Online]. Available: \url{https://ieeexplore.ieee.org/document/8823398}
\BIBentrySTDinterwordspacing

\bibitem{Wu2020ForensicTools}
\BIBentryALTinterwordspacing
T.~Wu, F.~Breitinger, and S.~O'Shaughnessy, ``Digital forensic tools: Recent advances and enhancing the status quo,'' \emph{Forensic Sci. Int.: Digital Investigation}, vol.~34, p. 300999, 2020. [Online]. Available: \url{https://doi.org/10.1016/j.fsidi.2020.300999}
\BIBentrySTDinterwordspacing

\bibitem{Wickramasekara2024PotentialLLM}
\BIBentryALTinterwordspacing
A.~Wickramasekara, F.~Breitinger, and M.~Scanlon, ``Exploring the potential of large language models for improving digital forensic investigation efficiency,'' 2024, arXiv:2402.19366. [Online]. Available: \url{https://arxiv.org/pdf/2402.19366}
\BIBentrySTDinterwordspacing

\bibitem{Scanlon2023ChatGPTForensics}
\BIBentryALTinterwordspacing
M.~Scanlon, B.~Nikkel, and Z.~Geradts, ``Digital forensic investigation in the age of chatgpt,'' \emph{Forensic Sci. Int.: Digital Investigation}, vol.~44, p. 301543, 2023. [Online]. Available: \url{https://doi.org/10.1016/j.fsidi.2023.301543}
\BIBentrySTDinterwordspacing

\bibitem{Xu2022PresentableEvidence}
\BIBentryALTinterwordspacing
W.~Xu and D.~Xu, ``Visualizing and reasoning about presentable digital forensic evidence with knowledge graphs,'' in \emph{Proc. 19th Int. Conf. Privacy, Security \& Trust (PST)}, 2022, pp. 1--10. [Online]. Available: \url{https://ieeexplore.ieee.org/document/9851972}
\BIBentrySTDinterwordspacing

\bibitem{Addai2024GraphEvidence}
D.~Addai, S.~Shaikh, E.~Xu, W.~Zhang, and W.~Xu, ``A graph-based approach for discovering evidence relationships across multiple devices in group crimes,'' in \emph{Proc. 24th IEEE Int. Conf. Software Quality, Reliability, and Security Companion (QRS-C)}, 2024, pp. 1312--1313.

\bibitem{Karim2024GraphTheory}
\BIBentryALTinterwordspacing
A.~Karim \emph{et~al.}, ``Forensic artifacts' analysis using graph theory,'' in \emph{CEUR Workshop Proc.}, 2024. [Online]. Available: \url{https://ceur-ws.org/Vol-3792/paper26.pdf}
\BIBentrySTDinterwordspacing

\bibitem{Xu2024AIForensics}
\BIBentryALTinterwordspacing
E.~Xu, W.~Zhang, and W.~Xu, ``Transforming digital forensics with large language models: Unlocking automation, insights, and justice,'' in \emph{Proc. ACM Int. Conf. Information and Knowledge Management (CIKM)}, 2024, pp. 1450--1462. [Online]. Available: \url{https://dl.acm.org/doi/pdf/10.1145/3627673.3679091}
\BIBentrySTDinterwordspacing

\bibitem{Pan2024LLM_KG_Roadmap}
\BIBentryALTinterwordspacing
S.~Pan, L.~Luo, Y.~Wang, C.~Chen, J.~Wang, and X.~Wu, ``Unifying large language models and knowledge graphs: A roadmap,'' \emph{IEEE Trans. Knowl. Data Eng.}, 2024. [Online]. Available: \url{https://ieeexplore.ieee.org/document/10387715}
\BIBentrySTDinterwordspacing

\bibitem{Wickramasekara2024HybridApproach}
\BIBentryALTinterwordspacing
A.~Wickramasekara, F.~Breitinger, and M.~Scanlon, ``Where is the potential for large language models in digital forensic investigations?'' in \emph{Proc. Digital Forensics Research Workshop (DFRWS EU)}, 2024, pp. 78--89. [Online]. Available: \url{https://dfrws.org/wp-content/uploads/2024/03/DFRWS_EU_2024_paper_4391.pdf}
\BIBentrySTDinterwordspacing

\bibitem{Talekar2024ForenSift}
\BIBentryALTinterwordspacing
S.~Talekar \emph{et~al.}, ``Forensift: Gen-ai powered integrated digital forensics and incident response platform using langchain framework,'' \emph{Int. J. Multidisciplinary Research (IJFMR)}, 2024. [Online]. Available: \url{https://www.ijfmr.com/research-paper.php?id=31692}
\BIBentrySTDinterwordspacing

\bibitem{OASIS2020STIX}
\BIBentryALTinterwordspacing
{OASIS Committee Specification}, ``Stix version 2.1. part 1: Stix core concepts,'' Online, March 2020, accessed: 2025-05-16. [Online]. Available: \url{https://docs.oasis-open.org/cti/stix/v2.1/stix-v2.1-part1-stix-core.html}
\BIBentrySTDinterwordspacing

\bibitem{Cellebrite2021CTF}
\BIBentryALTinterwordspacing
C.~2021, ``Overview of the 2021 cellebrite capture the flag event,'' 2021. [Online]. Available: \url{https://cellebrite.com/en/overview-of-the-2021-cellebrite-capture-the-flag-event/}
\BIBentrySTDinterwordspacing

\bibitem{Cellebrite2022CTF}
\BIBentryALTinterwordspacing
C.~2022, ``Digital forensics 101: The value of ‘capture the flag’ events,'' 2022. [Online]. Available: \url{https://cellebrite.com/en/blog/final-ctf-2022-round-up/}
\BIBentrySTDinterwordspacing

\end{thebibliography}
\nocite{*}

\end{document}